\documentclass[%
 reprint,
 amsmath,amssymb,
 aps,nofootinbib,   
]{revtex4-1}
\usepackage{bm}
\PassOptionsToPackage{linktocpage}{hyperref}
\usepackage[hyperindex,breaklinks]{hyperref}
\usepackage{enumitem}
\usepackage{slashed}
\usepackage{float}

\renewcommand{\theta}{\vartheta}

\usepackage{soul}

\usepackage{array}
\usepackage{mathtools}
\usepackage{xcolor}

\usepackage{etoolbox}
\smallskipamount
\begin{document}

\title{\texorpdfstring{Minimal $\bm{SO(10)}$}{} \textit{ante portas}: the importance of being effective}

\author{Anca Preda$^{1}$, Goran Senjanovi\'c$^{2,3}$ and Michael Zantedeschi$^{4,5}$}
\affiliation{%
$^1$
Department of Physics, Lund University, SE-223 62 Lund, Sweden}
\affiliation{
$^2$
Arnold Sommerfeld Center, Ludwig-Maximilians University, Munich, Germany
}%
\affiliation{%
$^3$
International Centre for Theoretical Physics, Trieste, Italy
}%
\affiliation{$^4$ Tsung-Dao Lee Institute (TDLI), 
Shanghai, China}
\affiliation{$^5$ School of Physics and Astronomy, Shanghai Jiao Tong University, 
Shanghai, China}

\begin{abstract}
The minimal $SO(10)$ grand unified theory, augmented by higher-dimensional operators, is based on the following Higgs representations: adjoint $45_{\rm H}$, spinor $16_{\rm H}$ and complex vector $10_{\rm H}$. It was recently realized that, as opposed to the conventional wisdom, any intermediate mass scale has to lie close to the grand unification one. This bears profound consequences for the scalar particles mass spectrum and the proton decay lifetime. 
A complete analysis of the parameter space shows the following interesting correlation under the requirement of perturbativity of the theory: A color octet below approximately $5\,\rm TeV$ would imply a lower bound on neutrino mass $m_{\nu}\gtrsim 0.2\,\rm eV$, soon to be probed, a scalar weak triplet below $500\,\rm GeV$ and a scalar weak doublet, color triplet below $3\,\rm TeV$. If one were to stick to no flavor cancellations in proton decay amplitudes, one would recover a recent result of all these states being accessible at collider energies. Moreover, the proton decay lifetime would then be tantalizingly close to the present experimental bound. 
\end{abstract}

\maketitle

\section {Introduction} 

By providing a rationale for charge quantization in nature and by unifying relevant particle interactions, grand unification leads to two profound predictions, the existence of magnetic monopoles and proton decay. The minimal such theory is based on $SU(5)$ gauge group and in its original form~\cite{Georgi:1974sy} it fails for three different reasons: gauge couplings do not unify, as seemed originally~\cite{Georgi:1974yf}, neutrino mass vanishes just as in the Standard Model (SM) and charged fermion masses are wrongly predicted. 

Remarkably enough, all three can be cured by the addition of $d = 5$ effective interactions~\cite{Senjanovic:2024uzn}, as long as one is willing to accept flavor rotations in proton decay. It was known almost at the outset that such terms cure bad charged fermion mass relations~\cite{Ellis:1979fg} and provide non-vanishing neutrino mass~\cite{Weinberg:1979sa}, and with time it was understood that they also help unification~\cite{Shafi:1983gz}. Still, for many years, it was claimed that this theory was ruled out because of too fast proton decay.

While it is reassuring that the minimal grand unified theory (GUT) is still valid, it is worth stressing that it does not unify the SM fermions and strictly speaking, it is tailor-made for massless neutrinos. Instead, it is the $SO(10)$ GUT~\cite{Georgi:1974my,Fritzsch:1974nn} that automatically does the job of unifying a generation of fermions in a spinor representation $16_{\rm F}$ and providing neutrino mass. The reason is deep and simple: while $SU(5)$ unifies the SM electroweak gauge group, $SO(10)$ contains its left-right (LR) symmetric extension~\cite{Mohapatra:1974gc,Senjanovic:1975rk,Senjanovic:1978ev}, which by default requires both the existence of the right-handed neutrino (RH) and thus that neutrino be massive. The LR symmetry is actually a finite gauge symmetry in the form of (generalised) charge conjugation~\cite{Slansky:1980gc, Slansky:1981yr} and moreover, the theory possesses the Pati-Salam quark-lepton (QL) symmetry~\cite{Pati:1974yy}.

Furthermore, through the seesaw 
mechanism~\cite{Minkowski:1977sc,Yanagida:1979as,Mohapatra:1979ia,Glashow:1979nm,GellMann:1980vs}, it offers a natural explanation of a tiny neutrino/electron mass ratio, and connects it to the near-maximum violation of parity in nature.

Lastly, gauge coupling unification seemed to have naturally been obtained~\cite{Shafi:1979qb,delAguila:1980qag,Rizzo:1981su}, precisely through the existence of an intermediate mass scale associated with the same LR symmetry in question. 

The situation with the $SO(10)$ depends on the choice of the Higgs sector, which follows two opposing roads: small representations that require higher-dimensional operators in order to reproduce a realistic fermion mass spectrum, but are easier to deal with, and large representations that are harder to handle but can work at the renormalisable level. The latter approach, if one sticks to the principle of minimality, offers hope for predicting fermion mass and mixing relations, important for proton decay branching ratios. While appealing in principle, it is hard, if possible at all\footnote{See recent works on the minimal renormalizable $SO(10)$ model that, ultimately, seems to be ruling it out in its entire parameter space~\cite{Graf:2016znk,Jarkovska:2023zwv,Susic:2024hgc}.}, in practice. The main con is the proliferation of physical scalar states that prevents a serious study of gauge coupling unification, and so in order to claim ``predictions'', one subscribes to the so-called extended survival principle which assumes - without any justification - that particle masses take the largest possible values consistent with symmetries in question~\cite{delAguila:1980qag,Mohapatra:1982aq}. 

The survival principle was first proposed in~\cite{Georgi:1979md} for the case of fermions, and both the original and extended versions have been typically used to deal with the unification of the gauge coupling. However, there is no physical reason for this principle to be correct, and so no way of trusting the resulting claims. After all, if one is willing to accept the fine-tuning of the SM Higgs mass, one can as well accept it for other scalars. Of course, one can assume the minimal fine-tuning principle, but again, there is no guarantee that it works. Strictly speaking, this is a wrong way of doing physics. After all, a study of a theory is a struggle to fix the parameter space from experiment, consistency and theoretical structure - not choosing the parameters aesthetically. 

The extended survival principle found its fertile application especially in $SO(10)$ for another important reason. It fitted beautifully with an existence of an intermediate LR symmetry breaking scale, tailor made for small neutrino mass, and it became the standard of the field.

Going back to the choice of Higgs representations, in the case of small ones, one must employ higher-dimensional operators, and so the theory suffers from the proliferation of couplings, which obscures the question of proton decay amplitudes. The crucial point, though, is that this program offers hope for a complete study of unification conditions since it allows for a treatment of threshold effects of the unknown scalar masses, and moreover it can dramatically restrict these parameters.

In this spirit, we have recently readdressed~\cite{Preda:2022izo} the question of gauge coupling unification for a minimal version of the theory with small Higgs representations, consisting of an adjoint $45_{\rm H}$, a spinor $16_{\rm H}$ and a complex vector $10_{\rm H}$. Under the common assumption of no judicious cancellations in proton decay branching ratios, we found - contrary to the usual prejudice of a desert in energies from the weak and the GUT scale - a number of new light scalar states potentially accessible at the LHC: scalar gluon, scalar $W,Z$ (already present in the minimal $SU(5)$) and an additional scalar quark doublet (a leptoquark).  

Our results would seem to manifestly invalidate the survival principle, that would imply these  light states to have masses at the GUT scale. 
At the same time, we found that proton lifetime would end up being below $10^{35} \,{\rm years}$, expected to be probed in the planned experiments. Thus, relatively short proton lifetime and the existence of light states at today's energies, would seem to go hand in hand, and would make the theory rather exciting from the experimental point of view.

However, the flavour freedom of proton decay obscures these apparent predictions - after all, as we said, even the minimal $SU(5)$ theory remains viable with higher-dimensional operators~\cite{Senjanovic:2024uzn}. Thus, it would seem that with a sufficient rotation of proton decay amplitudes, these scalar states can become heavy and effectively unreachable in any foreseeable future. 

However this conclusion is wrong. Namely, the key-operators ensuring unification in the minimal $SU(5)$ effective theory~\cite{Senjanovic:2024uzn} are the ones affecting the gauge kinetic term introduced by~\cite{Shafi:1983gz}, which correct non-trivially the gauge coupling unification condition. Simply, without them, unification fails. While in the minimal $SU(5)$ model such operators contribute at $d=5$, in the minimal $SO(10)$ theory we consider here they start contributing at dimension 6, with a much less significant impact. Therefore, the burden of ensuring unification is again on the scalar particle states, which, as a consequence, ought to be light even in this model. 

Although no hard prediction is found, an interesting scenario emerges: 
\textit{If the scalar color octet was accessible at the LHC, or the next hadron collider, neutrino mass would lie above $0.2\, \rm eV$, close to the near future reach of KATRIN~\cite{Katrin:2024tvg}. Moreover, a plethora of new particle states, including a scalar weak triplet and a scalar quark, would also be found at collider energies.
}

The rest of this paper is organized as follows. In the next Section we review the salient features of the $SO(10)$ theory, relevant for our study. We show there how the smallness of neutrino mass, as well as the $b-\tau$ mass ratio, force intermediate symmetry breaking scales to be large, close to the GUT one. This plays a crucial role in the unification constraints, suggesting new light states as discussed above. In Section III we give an in-depth analysis of the version of the theory based on the $45_{\rm H}$, $16_{\rm H}$ and a complex vector $10_{\rm H}$ Higgs sector. Finally, in the last Section we give the summary of our results and and offer an outlook for the future.

\section {$SO(10)$ theory with $45_{\rm H}$, $16_{\rm H}$ and complex $10_{\rm H}$} 

The minimal version of the theory (with small representations) besides three $16_{\rm F}$ spinors (fermion generations), contains, as we said, the following Higgs scalars
\begin{equation}\label{higgs}
45_{\rm H}; \,\,\,\,\,  16_{\rm H}; \,\,\,\,\,  10_{\rm H}\,,
\end{equation}
where we use the notation that specifies the representation content. The $45_{\rm H}$ field is an adjoint, antisymmetric representation which, together with the spinorial $16_{\rm H}$ Higgs field, serves to break  the GUT symmetry to the SM gauge symmetry. 
 Finally, a complex $10_{\rm H}$ vector is where the SM Higgs doublet resides, which then completes the breaking down to electromagnetic charge and color gauge invariance.

The truly minimal Higgs sector would use a real $10_{\rm H}$, but that would imply a single Yukawa coupling $16_{\rm F}10_{\rm H} 16_{\rm F}$ and a single vacuum expectation value, predicting all fermion masses being equal. In particular, top-bottom mass equality would be impossible to cure, since the higher-dimensional operators are not large enough. This is why  $10_{\rm H}$ must be complex, with two Yukawa couplings written schematically
\begin{equation}
\label{eq:d4yukawa}
   \left({\cal L}_{\rm Y}\right)^{d=4} = 16_{\rm F}16_{\rm F}(10_{\rm H} + 10_{\rm H}^*)\,,
\end{equation}
and two VEVs, which allows to split up from down quarks. 

While necessary, this is not sufficient. The SM doublets in $10_{\rm H}$ are singlets under Pati-Salam quark-lepton symmetry, which implies the same masses for down quarks and charged leptons, clearly wrong. One could add more Yukawa Higgs scalars, but that would imply either $120_{\rm H}$ or/and $126_{\rm H}$ representation, against the original premise of employing the small representations. Thus, one must augment the renormalisable Lagrangian with higher-dimensional operators. One has then the necessary part of Yukawa sector, schematically written
\begin{equation}\label{yukawa}
\begin{split}
\left({\cal L}_{\rm Y}\right)^{d \geq  5} &= 16_{\rm F} 16_{\rm F} \left\{  \left(10_{\rm H} + 10_{\rm H}^* \right) \frac {45_{\rm H}}{\Lambda} + \frac {16_{\rm H} 16_{\rm H}} {\Lambda}\right. \\&\qquad \left.+ \frac {16_{\rm H}^* 16_{\rm H}^*} {\Lambda}
+ \left(10_{\rm H} + 10_{\rm H}^* \right)\frac{45_{\rm H}^2}{\Lambda^2}+...\right\}\,,
\end{split}
\end{equation}
where $\Lambda$ denotes the cutoff of the theory. In order for this expansion to be perturbatively valid, we take hereafter $\Lambda \gtrsim 10 M_{\rm GUT}$. The reader should keep in mind that this schematic writing does not count the number of such Yukawa couplings. 

While \eqref{eq:d4yukawa} provides the mass for the top quark, the first and fourth term in \eqref{yukawa} correct the charged lepton-down quark mass equality. Naively, one could imagine that the first term suffices for this, but the problem is that $45_{\rm H}10_{\rm H}$ can amount only to $10_{\rm H}$ and $120_{\rm H}$ direct Yukawa terms. Now, $10_{\rm H}$ is responsible for wrong mass relations, and $120_{\rm H}$ couplings are anti-symmetric and thus insufficient for this task - hence the necessity for the fifth $d=6$ term. We are rushing ahead of ourselves, though, we discuss this carefully below once the symmetry breaking patterns are established.

Finally, the third term is also crucial since it sets the stage for the seesaw mechanism of neutrino mass, as we discuss below. Notice, that we only wrote down the necessary terms; one should keep in mind that there are other $d=5$ terms such as $16_{\rm F}^216_{\rm H}^2$ that could in principle contribute to light fermion masses (see \cite{Shukla:2024owy} for a recent discussion of such coupling). More on this in Subsec.~\ref{subsection:weakscalesymmetrybreaking}. \\

\subsection{Neutrino Mass: the Crux of it All} 
Without the fourth term in \eqref{yukawa}, neutrino, as other fermions, would get only the Dirac mass from the first two terms and its lightness would remain a mystery. The fourth term, however, provides the mass for the RH neutrino N (the $SU(5)$ singlet heavy lepton), on the order of
\begin{equation}\label{Nmass}
m_{\rm N} \simeq \frac {\langle 16_{\rm H} \rangle^2}{\Lambda}\,.
\end{equation}

There is also the two-loop radiatively induced N mass~\cite{Witten:1979nr}, 
$m_{\rm N} \lesssim (\alpha / \pi)^2 \langle 16_{\rm H} \rangle^2 / M_{\rm GUT}$. Now, there is an upper limit on the cutoff, naively the Planck scale, but in reality scaled down by the number of real physical states (n) of the theory $\Lambda \lesssim M_{\rm Pl} / \sqrt n$~\cite{Dvali:2007hz,Dvali:2007wp}. Since $n \gtrsim 100$ in this theory, and, naively, $M_{\rm GUT} \gtrsim 10^{15} {\rm GeV}$ for the sake of proton's stability, one has $\Lambda \lesssim 100\, M_{\rm GUT}$. This implies that the dominant contribution to $m_N$ comes from \eqref{Nmass}.

Once the SM symmetry breaking is turned on, implying non-vanishing Dirac mass term $M_D$, light neutrinos get their masses through the seesaw mechanism
\begin{equation}\label{seesaw}
M_\nu \simeq - M_D^T \frac{1}{M_N} M_D\,.
\end{equation}

Due to the presence of higher-dimensional operators, we lose control over $M_D$ and $M_N$, so it would appear that no predictions could be made. The point, however, is that the top quark gets the mass from the leading $d=4$ term in \eqref{yukawa}, implying then 
\begin{equation}\label{thirddirac}
m_{\rm D3} \simeq m_{\rm t}\,,
\end{equation}
where $m_{\rm D3}$ is the third generation neutrino Dirac mass term. 
This innocent looking relation turns out to be the crux to it all, since
in turn, from \eqref{Nmass} and \eqref{thirddirac}, one has a potentially too large neutrino mass
\begin{equation}
\begin{split}\label{nueffect}
&\frac{m_{\rm \nu}}{\rm eV} \simeq  \frac {(m_{\rm 3 D})^2} {{\rm eV}\,m_{\rm N}}  \simeq \frac {m_{\rm t}^2 \Lambda} {\rm eV\,\langle 16_{\rm H}\rangle^2} \simeq \\
	&\qquad \simeq\left(\frac{m_{\rm t}} {100 \rm {GeV}} \frac{ 10^{14}\rm Gev}{\langle 16_{\rm H} \rangle}\right)^2  \left(\frac{\Lambda}{ 10^{15}\rm GeV} \right),
 \end{split}
\end{equation}
where we normalize the scales in question for convenience - see below. 

\begin{figure*}[th!]
\centering
    \includegraphics[width=.9\textwidth]{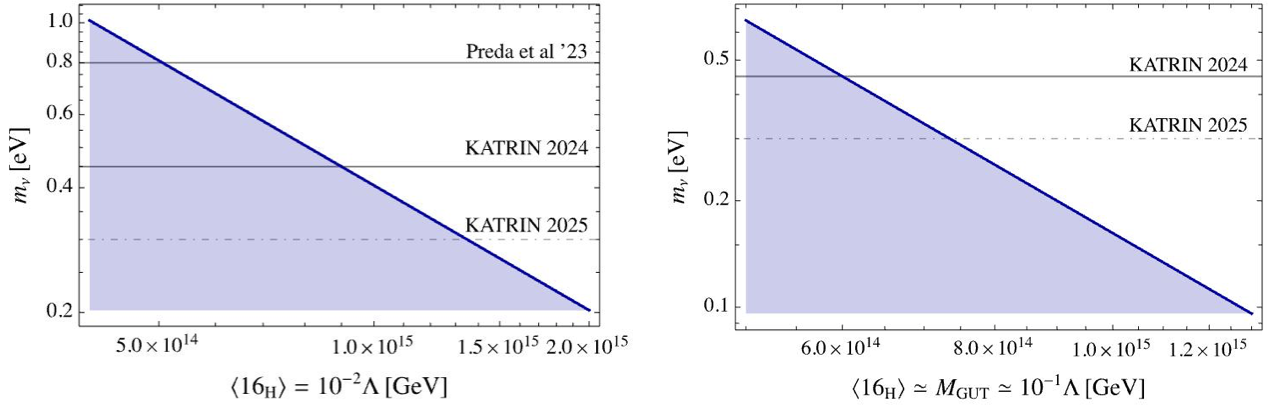}
    \caption{Comparison of neutrino mass scale for two-step breaking (left panel) and single-step breaking (right panel). Filled areas are excluded.
    The horizontal lines show present and future bounds on neutrino mass from KATRIN~\cite{Katrin:2024tvg}.
  }
    \label{fig:numass}
\end{figure*}

The smallness of neutrino mass sets the stage then for the main predictions of the theory. The way unification was imagined to work all these years - in the usual picture of the desert in energies from $M_{\rm W}$ to the intermediate scale $M_{\rm I}$ - was simple, albeit wrong. The problem of gauge coupling unification of SM forces is the fact that $\alpha_1$ hits $\alpha_2$ too early, much below the lower limit on the GUT scale from proton longevity - and the only way to slow down the rise of $\alpha_1$ is to embed the $U(1)$ gauge group into a non-Abelian subgroup. 

This seemed tailor made for the $SO(10)$ theory which possess a LR symmetric subgroup $SU(2)_{\rm L} \times SU(2)_{\rm R} \times U(1)_{\rm B-L} \times SU(3)_{\rm C}$ - with $M_{\rm I} = M_{\rm R} \simeq 10^{12}\, {\rm GeV}$, gauge coupling unification works apparently perfectly. This has been the conventional wisdom for decades now, but as \eqref{nueffect} clearly shows, it fails in this minimal version of the theory with small representations - it produces too big neutrino mass. 

Notice that Eq.~\ref{nueffect} is valid at the GUT scale - in fact, we normalize top mass with its value obtained from the running (see e.g., Table 3 in~\cite{Babu:2016bmy}). To compare neutrino mass with existing limits, we must take into account running effects, which roughly double its value at high energies~\cite{Ray:2010rz}. 
We have an indirect GERDA limit on neutrino Majorana mass from neutrinoless double beta decay $m_\nu \lesssim 0.2\, {\rm eV}$~\cite{GERDA:2020xhi}, but is obscured by flavor mixings - unlike the KATRIN direct limit  $m_\nu \lesssim \,0.45 \,{\rm eV}$~\cite{Katrin:2024tvg} from the endpoint of beta decay. This limit is expected to be pushed to $0.3\,\rm eV$ by the end of 2025 and lower further in the incoming years. 

Comparable bounds on neutrino mass, are obtained from cosmological considerations, $m_{\nu}\lesssim \, 0.1\, \rm eV$~\cite{Wang:2024hen}. Being indirect, however, these bounds, are less settled and reliable\footnote{We thank Uros Seljak and Zvonimir Vlah for a discussion on this issue.}.

To ease the reader's pain, we report the constraints on the $\langle 16_{\rm H}\rangle$ from the minimal possible neutrino mass in Fig.~\ref{fig:numass}. 
The blue shaded area corresponds to excluded regions. For clarity, because $\langle 16_{\rm H}\rangle$ is not necessarily exactly degenerate with $M_{\rm GUT}$, we report on the left side the case in which an hierarchy between the two scales is introduced, i.e., $\Lambda \sim 10\, M_{\rm GUT}\sim 100\,\langle16_{\rm H}\rangle $. The top horizontal line in the Figure corresponds to the two step breaking setup we considered in~\cite{Preda:2022izo}, which assumed an upper limit on neutrino mass around $0.8 \,\rm eV$. Current improvements on the neutrino mass constraints make the prediction of new light particle states of \cite{Preda:2022izo} even stronger. 

In the right panel, the exactly degenerate single-step breaking is shown instead. In this ideal scenario, the $\langle 16_{\rm H}\rangle$ scale is the least constrained by neutrino mass. For example, a neutrino mass below $0.1\,$eV, requires $\langle 16_{\rm H}\rangle$ above $\gtrsim 5\cdot 10^{14}\,\rm GeV$. Needless to say, other constraints of the theory, such as Yukawa viability as well as gauge coupling unification consistency need to be taken into account when reading the Figure, as we will show below.

The end result is clear: the intermediate scale $\langle 16_{\rm H}\rangle$ must be huge, basically as large as the GUT scale. Moreover, the same requirement emerges if the vacuum direction of $\langle 45_{\rm H}\rangle$ is preserving $SU(4)_{\rm C}$ symmetry. As we will see below, this follows from the $b-\tau$ mass ratio, which, in the absence of higher-dimensional operators, is predicted wrongly.

What about unification scale and proton decay?
Normally, one argues that the bound $\tau_{\rm p} \gtrsim 10^{34} \,{\rm yr}$~\cite{Super-Kamiokande:2020wjk} implies $M_{\rm GUT} \gtrsim 4 \cdot 10^{15}\, {\rm GeV}$. This is true under the condition that there are no judicial cancellations in proton decay amplitudes due to flavour conspiracies. In this limit, we~\cite{Preda:2022izo} found that consistency of the theory requires new particle thresholds at current collider energies. 

However, if one is to exploit the full freedom allowed by higher-dimensional Yukawa operators, proton decay amplitude can be suppressed by the tiny CKM mixing angle $\theta_{13}$~\cite{Nandi:1982ew}, resulting in a unification scale as low as roughly $10^{14}\rm GeV$~\cite{Dorsner:2004xa,Senjanovic:2024uzn}, compatible with proton lifetime. Notice that this is precisely the reference value we adopted in \eqref{nueffect}.

Before turning to a full analysis of the gauge unification constraints, and finding the range of the particle thresholds, we need to address the symmetry breaking patterns of the model. 

\subsection{GUT-Scale Symmetry Breaking}  
Let us start first with the adjoint $45_{\rm H}$ Higgs multiplet, responsible for the GUT scale symmetry breaking. Notice that, alternatively, $16_{\rm H}$ would break the $SO(10)$ symmetry down to $SU(5)$ (see below). 

The important question that must be faced is thus the pattern of symmetry breaking down to the SM gauge group. At the tree level, the answer is unique and unphysical: the combination of $\langle 45_{\rm H} \rangle$ and $\langle 16_{\rm H} \rangle$ leaves all of the 
$\rm {SU(5)}$ (and more) symmetry unbroken. However, a simple inclusion of radiative corrections \textit{à la} Coleman-Weinberg solves the problem and allows for two different channels of intermediate symmetry breaking~\cite{Bertolini:2009es}: $SU(2)_{\rm L} \times SU(2)_{\rm R} \times U(1)_{\rm B-L} \times SU(3)_{\rm C}$ (LR) or $SU(2)_{\rm L} \times U(1)_{\rm R} \times SU(4)_{\rm C}$ (QL). 
In other words, once the loops are included, we have physically interesting possibilities of left-right and quark-lepton symmetry, which have both played an important role in the BSM physics. Still, this leads to the presence of sum-rules between the masses of the particles in the spectrum.

With the inclusion of higher-dimensional operators, one may wonder first whether radiative corrections are needed, and second, whether one may have both LR and QL symetries following from $\langle 45_{\rm H}\rangle$ alone. After all, there are enough scales and couplings with the inclusion of $d=5$ and $d=6$ terms, to achieve this, at least in principle. When the dust settles, it turns out that the answer to the first question is positive, however the second possibility is still impossible. 

To see this, consider the following effective potential for $45_{\rm H}$ including operators up to $d=6$ terms 
\begin{equation}
\label{eq:potential45higherd}
\begin{split}
V=-\frac{1}{2} m^2  {\rm Tr} \left(45_{\rm H}^2 \right)+\frac{1}{4}  {\lambda}_1  {\rm Tr} \left(45_{\rm H}^2 \right)^2+\frac{1}{2}
   {\lambda }_2
    {\rm Tr} \left(45_{\rm H}^4 \right)\\
    +\frac{ {\rm Tr} \left(45_{\rm H}^4 \right)
    {\rm Tr} \left(45_{\rm H}^2 \right)}{4\,  {\mu_1^2}}+\frac{ {\rm Tr} \left(45_{\rm H}^2 \right)^3}{8\,  {\mu_2^2}}+\frac{ {\rm Tr} \left(45_{\rm H}^6 \right)}{2\,  {\mu_3^2}}\,.\quad
    \end{split}
    \end{equation}
The vacuum is given by 
 \begin{equation}
 \label{eq:45ansatzvLRQL}
 	\langle 45_{\rm H} \rangle = i \, \sigma_2 \otimes {\rm{diag}}(v_{\rm LR},v_{\rm LR},v_{\rm QL},v_{\rm QL},v_{\rm QL})\,,
 \end{equation}
where we denote by LR and QL the symmetries broken by the VEVs. At tree-level, the potential \eqref{eq:potential45higherd} allows only for the case $|v_{\rm LR}|=|v_{\rm QL}|$ which is $SU(5)\times U(1)$ preserving. A natural question is whether the additional $d=6$ terms in the second line of   \eqref{eq:potential45higherd} allows for more general symmetry breaking patterns. As already stated above, while it is possible to have them separately non-zero, the case where both $v_{\rm LR}$ and $v_{\rm QL}$ are non-vanishing is not viable.

   To see this, we evaluate the masses of two neutral states $B,D$ (for a more complete discussion see Appendix~\ref{app:masses})
 \begin{equation}
 \label{eq:45plusflucts}
 \begin{split}
 		\langle45_{\rm H}\rangle  &= i \, \sigma_2 \otimes {\rm{diag}}\left(v_{\rm LR}+B ,v_{\rm LR}-B\right.,\\
   &\qquad\qquad  \left. v_{\rm QL}+D,v_{\rm QL}-D,v_{\rm QL}\right)\,,\\
   m_B^2 &=24v_{\rm LR}^2 \left({v_{\rm LR}}^2- {v_{\rm QL}}^2\right)/{\mu_3}^2,\\
 	m_D^2 &=24 {v_{\rm QL}}^2 \left({v_{\rm QL}}^2-{v_{\rm LR}}^2\right)/{\mu_3}^2.
   \end{split}
 \end{equation}
 Clearly, one of the two states is tachyonic, which therefore excludes the possibility of a simultaneous breaking of LR and QL symmetries. It is interesting that the $d=4$ result persists, even with the inclusion of terms up to $d=6$. There could be a simple mathematical or physical reason behind this, but we are unaware of it. 
For completeness, we report the spectrum of the two respective breaking patterns in Appendix~\ref{QL_LR_mass}. 

In other words, the outcome of the study of the symmetry breaking implies just the following two possibilities 
\begin{equation}
 \begin{split}
&\langle 45_{\rm H} \rangle^{\rm LR} = v_{\rm GUT}\,i\, \sigma_2\, \otimes {\rm diag} (0, 0,1, 1, 1) \,,  \\
&\langle 45_{\rm H} \rangle^{\rm QL} = v_{\rm GUT}\, i\,\sigma_2\, \otimes {\rm diag} (1, 1,0, 0, 0)\,,\label{vev}
\end{split}
\end{equation}
for the LR and QL preserving cases, respectively. 

What happens at the next stage when $\langle 16_{\rm H} \rangle $ is turned on? Since the $16_{\rm H}$ has the same quantum numbers of the fermions in $16_{\rm F}$, the only neutral SM singlet field is the scalar RH neutrino. Once it gets the VEV, it obviously leaves all of the $SU(5)$ symmetry, as mentioned above. On the other hand, $\langle 16_{\rm H} \rangle$ breaks both $SU(2)_{\rm R}$ and $SU(4)_{\rm C}$ symmetries and does completes the breaking down 
to the SM symmetry group. 

There is more to it though. Once $\langle 16_{\rm H} \rangle\neq 0$ is turned on, it could affect the GUT stage of symmetry breaking [c.f. \eqref{vev}]. This follows from the tadpole interaction for the adjoint Higgs
\begin{equation}
\label{eq:tadpole}
    16_{\rm H}^*\,45_{\rm H}\,16_{\rm H}\,,
\end{equation}
which in turn might induce the missing VEVs (i.e., the zeroes) in \eqref{vev} on the order of $\langle 16_{\rm H}\rangle^2/v_{\rm GUT}$. 
This actually plays an important role for quark-lepton mass relations 
in the QL preserving channel as we show in the next Subsection. In fact,
besides the neutrino mass considerations above, this is yet another phenomenological reason for $\langle 16_{\rm H}\rangle$ to lie at the $v_{\rm GUT}$ scale. The importance of this cannot be overemphasised, since one is basically pushed to the single step breaking, which is known to be inconsistent - unless there are new light states populating the desert. And, as we said repeatedly, this is the essence of our work. 

\begin{figure}[]
    \centering
    \includegraphics[width=0.48\textwidth]{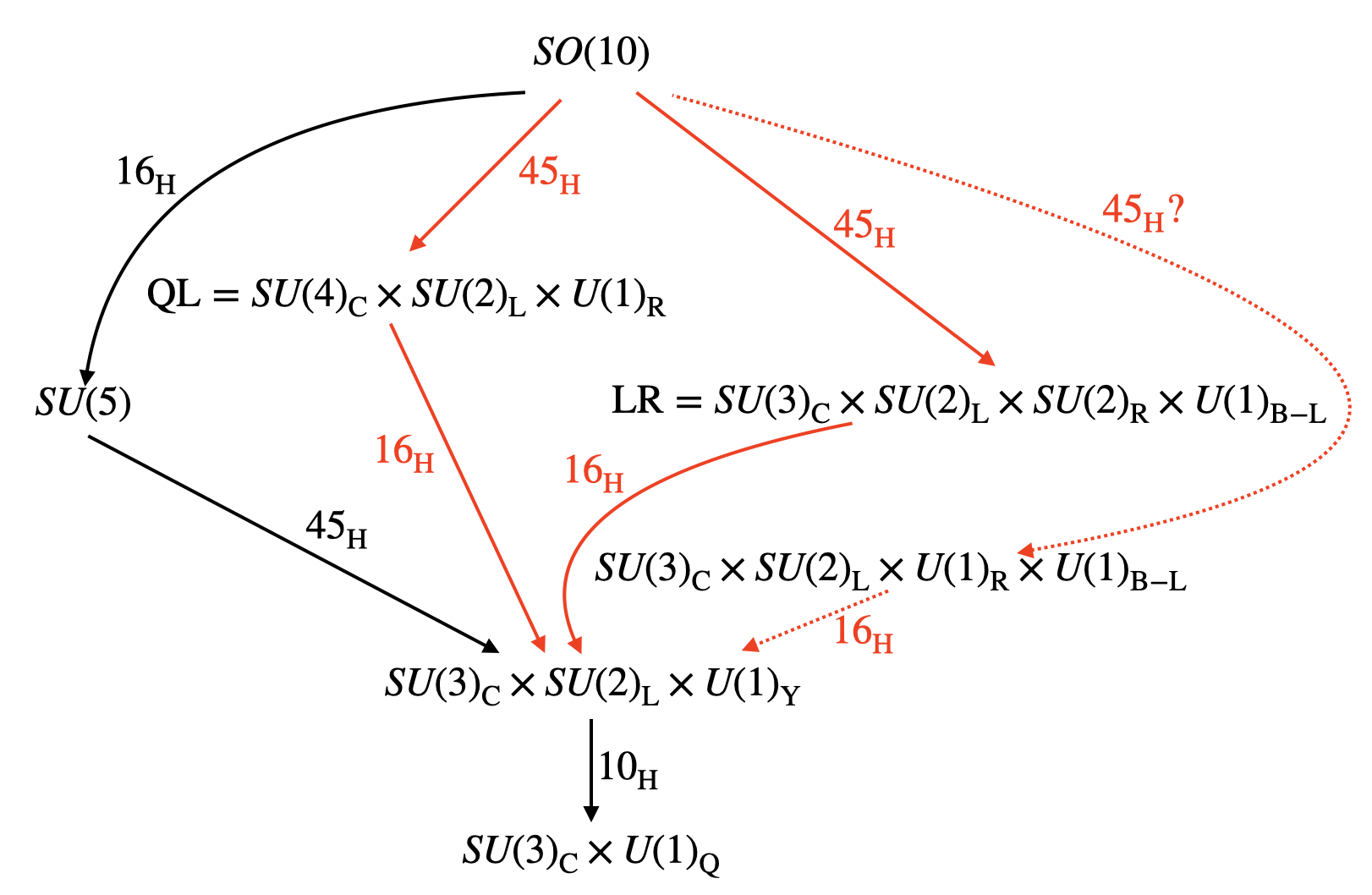}
    \caption{Possible breaking patterns. The black one corresponds to the one in which $\langle 16_{\rm H}\rangle$ determines the unification scale. As shown in Sec.~\ref{ref:unification} this direction gauge coupling unification forbids it. The red breaking pattern is the one allowed by the inclusion of higher-dimensional operators. At the level of our analysis, it is unclear wether $45_{\rm H}$ can break directly to $SU(3)_{\rm C}\times SU(2)_{\rm L} \times U(1)_{\rm R}\times U(1)_{\rm B-L}$.}
    \label{fig:breakingpatterns}
\end{figure}

Before moving to the discussion of the Yukawa couplings, we summarize in Fig.~\ref{fig:breakingpatterns} the possible breaking patterns. 

\subsection{Weak-Scale Symmetry Breaking} 
\label{subsection:weakscalesymmetrybreaking}
At first glance, here the situation is trivial since the light Higgs resides in $10_{\rm H}$ and thus the symmetry breaking apparently mimics the Standard Model or the $SU(5)$ one via fundamental $5_{\rm H}$. However the situation is more subtle: from the coupling 
\begin{equation}
    16_{\rm H}  10_{\rm H} 16_{\rm H}\,,
\end{equation}
in presence of the large $\langle 16_{\rm H}\rangle\neq 0$ in the direction of the scalar analog of the right-handed neutrino, through $\langle 10_{\rm H}\rangle$, also the scalar analog of the left handed neutrino gets an induced VEV $\langle 16_{\rm H}\rangle_{\rm light}$. 

Could $\langle 16_{\rm H}\rangle_{\rm light}$ dominate over $\langle 10_{\rm H}\rangle$? The answer is no, since this would imply a too large top quark Yukawa coupling. Namely, the only way $\langle 16_{\rm H}\rangle_{\rm light}$ can generate top quark mass is via the third term of the effective Yukawa interaction in \eqref{yukawa}, leading to
\begin{equation}
\label{mumd16}
   \frac{1}{\Lambda} 16_{\rm F} 16_{\rm F} \langle 16_{\rm H}^*\rangle \langle 16_{\rm H}^*\rangle_{\rm light}\,.
\end{equation}
This is equivalent to induced Yukawa couplings for the up quarks of order, at most, $M_{\rm GUT}M_W/\Lambda\,$. While this may be relevant for the first two generations, it is clearly insufficient for the top quark. In short, one can rest assured that the $d=4$ Yukawa coupling in~\eqref{eq:d4yukawa} generates the top quark mass and the third generation neutrino Dirac mass, implying~\eqref{thirddirac}. The reader should notice that \eqref{mumd16} gives only mass to up quarks and neutrinos. 

This, \textit{per se}, still does not prove that the SM Higgs does not lie predominantly in $16_{\rm H}$. After all, a small $\langle 10_{\rm H}\rangle$ could be compensated by a large $y_t$ in~\eqref{eq:d4yukawa}. The key-point is that perturbativity argument in grand unified theories tells us that the SM Yukawa top lies very close to its upper limit~\cite{Cabibbo:1979ay}. In other words, we have the usually imagined situation, where most of the SM Higgs belongs to $10_{\rm H}$, which significantly simplifies the analysis that follows. 

\subsection{Constraints from Yukawa Sector}
A natural question is whether the Yukawa couplings in (\ref{eq:d4yukawa}) and (\ref{yukawa}) can properly account for down quark and charged lepton masses. As already mentioned, at the renormalizable level two real $10_{\rm H}$ ensure the splitting between top and bottom mass. Still, one would still have 
$m_d=m_e$, which is clearly wrong. The first generation has tiny Yukawa couplings and, therefore, dimensions $d\geq 6$ operators can obviously correct the above wrong mass relation. In the case of second and third generations only the operators outlined in \eqref{yukawa} can address the issue and, \textit{a priori}, it is unclear whether they manage this. Specifically, we should check that those operators can reproduce the Yukawa couplings of the second and third generation without violating the requirement of perturbative validity i.e., $M_{\rm GUT}\lesssim 10^{-1}\Lambda$.

In complete analogy with Eq.~\eqref{mumd16}, the second term in~\eqref{yukawa} leads to induced down quark and charged lepton masses, via 
\begin{equation}
\label{memd16}
    \frac{1}{\Lambda} 16_{\rm F} 16_{\rm F} \langle 16_{\rm H}\rangle \langle 16_{\rm H}\rangle_{\rm light}\,.
\end{equation}
Since both $\langle 16_{\rm H}\rangle$ and $\langle 16_{\rm H}\rangle_{\rm light}$ manifestly break $SU(4)_{\rm C}$ symmetry, one would be tempted to conclude that~\eqref{memd16} modifies the tree-level relation $m_e = m_d$. However, both of these VEVs preserve an accidental symmetry between the down quarks and charge leptons. Therefore, it is the first and last term in \eqref{yukawa} that can do the job, since $\langle 45_{\rm H}\rangle$ (just as $\langle 24_{\rm H}\rangle$ in $SU(5)$) manifestly breaks the down quark-charge lepton symmetry. 

Before proceeding with the analysis, we report here the Yukawa couplings at the GUT scale~\cite{Babu:2016bmy}
\begin{equation}
\begin{split}
    &y_b= 0.57\cdot 10^{-2}\,,\quad
    y_s=1.24\cdot 10^{-4}\,,\\
    &y_{\tau}=0.97\cdot 10^{-2}\,,\quad y_{\mu}=5.7\cdot 10^{-4}\,.
\end{split}
\end{equation}

\paragraph*{\textbf{QL preserving pattern.}}
The issue is further complicated in case the symmetry breaking direction is QL preserving, i.e., $ \langle 45_{\rm H}\rangle^{\rm LR} = 0  \neq \langle 45_{\rm H}\rangle^{\rm QL} $. Notice that, due to the unbroken QL symmetry, no combination of the form $\langle 10_{\rm H}\rangle \, (\langle45_{\rm H} \rangle^{\rm QL})^n$ can correct the relation between down quark and charged leptons. The QL symmetry is broken, on the other hand, by $\langle 16_{\rm H}\rangle \neq 0$, which, through $\langle 16_{\rm H}\rangle_{\rm light}$, could in principle do the job. However, the operator in $\eqref{mumd16}$ contributed only to neutrino and up quark masses.  
Therefore, 
the only phenomenological viable option is for $45_{\rm H}$ to break the QL symmetry through the induced VEV via \eqref{eq:tadpole}.

As mentioned below Eq.~\eqref{yukawa}, the two relevant scalar operators to address the above Yukawas are $10_{\rm H}45_{\rm H}$ and $10_{\rm H} 45_{\rm H}^2$ amounting to an effective $120_{\rm H}$ and $126_{\rm H}$ direct Yukawa couplings. The former corresponds to an anti-symmetric contribution, while the latter is symmetric. Their amplitudes are estimated to be
\begin{equation}
\label{eq:yukamplitude}
\begin{split}
   &\frac{ \langle10_{\rm H}45_{\rm H}\rangle}{\Lambda}\supset \frac{\langle10_{\rm H}\rangle\langle 45_{\rm H}\rangle^{\rm LR}}{\Lambda}\lesssim  \frac{M_W\,\langle 16_{\rm H}\rangle^2}{M_{\rm GUT}\,\Lambda}\,,\\
   &\frac{\langle10_{\rm H}45_{\rm H}^2\rangle}{\Lambda^2} \supset \frac{\langle10_{\rm H}\rangle\langle 45_{\rm H}\rangle^{\rm QL}\langle 45_{\rm H}\rangle^{\rm LR}}{\Lambda^2}\lesssim  \frac{M_W\,\langle 16_{\rm H}\rangle^2}{\Lambda^2}\,.\qquad
\end{split}
\end{equation}
One choice elucidating our findings for the second and third generation Yukawas is given by
\begin{equation}
\label{eq:yukawamatrix}
\begin{split}
    &Y_e=\begin{pmatrix}
y_1 &\epsilon_1 + \epsilon_2\\
-\epsilon_1 + \epsilon_2 &y_2
\end{pmatrix}\,,\\
&Y_d=\begin{pmatrix}
y_1 &-3(\epsilon_1 + \epsilon_2)\\
-3(-\epsilon_1 + \epsilon_2) &y_2
\end{pmatrix}\,.
\end{split}
\end{equation}
where $y_{1,2}$ denote the couplings of $10_{\rm H}$ and  $10_{\rm H}^*$ respectively (the $SU(4)_c$ symmetry of $\langle 10_{\rm H} \rangle$ then ensures $Y_e = Y_d$); $\epsilon_1$ denotes the antisymmetric contribution from $10_{\rm H}45_{\rm H}$, and $\epsilon_2$ the one from $10_{\rm H}45_{\rm H}^2$. One can easily check that more general forms of the Yukawa matrix \eqref{eq:yukawamatrix} do not alter the qualitative outcome of our statement.

Some useful relations that follow from the above ansatz are
\begin{equation}
\label{eq:systemsolution}
\begin{split}
    &{\rm Tr}\left(Y_{d}^T\,Y_d - Y_e^T\,Y_e\right)/16 = -\epsilon_1^2-\epsilon_2^2\simeq -3.87\cdot 10^{-6}\,,\\
    & {\rm Tr}\left(9\,Y_{d}^Y\,Y_d - Y_e^T\,Y_e\right)/8 = y_1^2 + y_2^2\simeq 2.48 \cdot 10^{-5}\,,\\
    &\left({\rm Det}\left(Y_d^T\,Y_d\right)-{\rm Det}\left(Y_e^T\,Y_e\right)\right)/16=\\
    &\quad \,\,\,\left(y_1y_2 + 5\, \epsilon_1^2 -5\,\epsilon_2^2\right)(\epsilon_2^2-\epsilon_1^2)\simeq -1.88\cdot 10^{-12}\,,\\
    &{\rm Det}\left(Y_d^T\,Y_d\right)-{\rm Det}\left(Y_e^T\,Y_e\right)/9=\\
    &\quad \quad8\, y_1^2\, y_2^2/9- 8\left(\epsilon_1^2-\epsilon_2^2 \right)^2\simeq - 2.90 \cdot 10^{-12}\,,
    \end{split}
\end{equation}
which admit solutions for $|\epsilon_1|\simeq |\epsilon_2| \sim 10^{-3}$, and $|y_1|\sim 2\cdot 10^{-5}\ll |y_2|\sim  5\cdot 10^{-3}$. 

Therefore, Eq.~\eqref{eq:yukamplitude}, under the perturbative requirement $M_{\rm GUT}= {\rm max}\left(\langle 45_{\rm H}\rangle,\langle 16_{\rm H}\rangle\right)\lesssim 10 \Lambda$, together with $\epsilon_2 \sim 10^{-3}$, implies a lower limit on $\langle 16_{\rm H}\rangle$
\begin{equation}
    \langle 16_{\rm H}\rangle\gtrsim  \langle 45_{\rm H}\rangle^{\rm QL}/3\,,
\end{equation}
pushing the theory into single step breaking. 

\vspace{.3cm}

\paragraph*{\textbf{LR preserving pattern.}}
In the case $ \langle 45_{\rm H}\rangle^{\rm QL} = 0  \neq \langle 45_{\rm H}\rangle^{\rm LR}$, the Yukawa's conditions in \eqref{eq:systemsolution} are easily fulfilled, without the need of a large induced LR breaking VEV in $45_{\rm H}$ from $16_{\rm H}$. Therefore, no constraints in this case emerge on the $\langle 16_{\rm H}\rangle$. It is neutrino mass considerations that pushes it, as we already discussed, close to $M_{\rm GUT}$.

\section{Unification} \label{ref:unification}
We now turn to the main subject of our work, the constraints that arise from the demand for unification. 

In order to be general in our gauge coupling unification analysis, we ought to take into account the impact of additional higher-dimensional operators, affecting the gauge coupling kinetic term
\begin{equation}\label{eq:gaugekinetic}
    \begin{split}
        &{\rm Tr}\left(F_{\mu\nu}F^{\mu\nu} \langle 45_{\rm H}\rangle  \right)/\Lambda ={ 0}\,, \\
        &{\rm Tr}\left(F_{\mu\nu} \langle 45_{\rm H}\rangle  \right){\rm Tr}\left(F^{\mu\nu} \langle 45_{\rm H}\rangle  \right)/\Lambda^2\rightarrow \text{irrelevant}\,,\\
        &{\rm Tr}\left(F_{\mu\nu}F^{\mu\nu} \langle 16_{\rm H}^*\rangle\langle 16_{\rm H}\rangle  \right)/\Lambda^2\rightarrow \text{irrelevant}\,,\\
        &{\rm Tr}\left(F_{\mu\nu}F^{\mu\nu} \langle 45_{\rm H}^2\rangle  \right)/\Lambda^2\,,
    \end{split}
\end{equation}
where the first line denotes the single $d=5$ operator, while the last $3$ lines represent all possible $d=6$ invariants. Higher-dimensional invariants are negligible in the unification analysis. 

The first term vanishes due to the asymmetry of the adjoint representation. The second term, while non-vanishing, is clearly not differentiating between the gauge boson submultiplets in terms of SM symmetry group. Similarly, the third term also leads to no correction to the gauge coupling unification, since it preserves the full $SU(5)$ symmetry.

Therefore, we arrive to the conclusion that only the last operator in \eqref{eq:gaugekinetic} corrects the gauge coupling unification conditions. This is a fortuitous simplification, that significantly eases the gauge coupling unification analysis. Explicitly, the gauge couplings at the GUT scale, obey the condition
    \begin{equation}
\label{eq:gaugeunif}
	\begin{split}
	\alpha_3+\delta\alpha_3 = \alpha_2+\delta\alpha_2 = \alpha_1+\delta\alpha_1
	\end{split}
\end{equation}
where $\delta \alpha_i$ coefficients are
\begin{equation}
    \label{eq:deltaalphai}
    \delta \alpha_1 = \frac{1}{\Lambda^2}\left(\frac{2}{5}v_{\rm LR}^2 + \frac{3}{5}v_{\rm QL}^2 \right),\,
    \delta \alpha_2=\frac{v_{\rm LR}^2}{\Lambda^2},\,
    \delta \alpha_3=\frac{v_{\rm QL}^2}{\Lambda^2}\,.
\end{equation}
We refer the reader to Appendix~\ref{app:deltaalphai} for a detailed derivation. The above result generalizes the findings of~\cite{Preda:2022izo} which focused on the two step breaking, therefore assuming either $v_{\rm QL}$ or $v_{\rm LR}$ to be negligible. Notice that in the limit $v_{\rm LR}= \pm v_{\rm QL}$, Standard Model gauge coupling are corrected equally, resulting in an overall factor, compatibly with the fact that, in this limit, $SU(5)$ symmetry is enforced. 
In what follows we make the perturbative assumption $\delta \alpha_1 \lesssim 10^{-2}$, compatible with a cutoff requirement $\Lambda \gtrsim 10M_{\rm GUT}$.

An immediate consequence of \eqref{eq:deltaalphai}, is that it is not possible to obtain gauge coupling unification for $\langle 16_{\rm H}\rangle \gg \langle 45_{\rm H}\rangle$.
The reason for this lies in the failure of gauge coupling unification. In fact, one of the main ingredient necessary to ensure unification in the minimal $SU(5)$ effective theory~\cite{Senjanovic:2024uzn}, is the inclusion of non-negligible $d=5$ gauge coupling operator. In their absence, unification is never obtained - a known shortcoming of the renormalizable Georgi-Glashow model. 

Differently from  the minimal $SU(5)$ spectrum, a higher number of new particle thresholds can aid unification. However, a potential hierarchy between the high $\langle 16_{\rm H}\rangle$ scale and a low $\langle 45_{\rm H}\rangle$, is impossible: simply, the $SU(5)$ scale would need to be too low. This is due to the fact that very few particle states, if light, would help unification. These consist of the weak triplet and color octet, present also in the minimal $SU(5)$ model, the scalar quark with SM charges $(3_{\rm C},2_{\rm L},1/6_{\rm Y})$, which, due to the non-vanishing hypercharge, and color charge, aids unification only marginally, and two Higgs doublets in complex $10_{\rm H}$ and $16_{\rm H}$. These particle states alone are insufficient to ensure a sufficiently low unification scale. 
Therefore, one is pushed into single step breaking also in this case.

\subsection{Unification Diagnostics}

To start with, we need to know the complete particle spectrum, which contains the following three sectors:

\textit{(i) Fermion sector. } It consists of the usual three SM generations of quarks and leptons, augmented by SM singlet RH neutrinos N.

\renewcommand{\arraystretch}{1.2}
\begin{table}[h!]
\centering
\begin{tabular}{cccc|c}
\hline 
 $4_C\,2_L\,2_R $
& $4_C\,2_L\,1_R $
& $3_c\,2_L\,2_R\,1_{BL} $
& $3_c\,2_L\,1_Y $
&fermions
\\
\hline
 $\left({ 4,2,1} \right)$
& $\left(4,2,0 \right)$
& $\left(3,2,1,1/6 \right)$
& $\left(3,2,1/6 \right)$
&$q_{\rm L}$
\\
\null
& 
& $\left(1,2,1,-1/2 \right)$
& $\left(1,2,-1/2 \right)$
& $l_{\rm L}$
\\
$\left({ \overline{4},1,2} \right)$
& $\left(\overline{4},1,1/2\right)$
& $\left(\overline{3},1,2,-1/6 \right)$
& $\left(\overline{3},1,1/3 \right)$
& $d_{\rm R}$
\\

& 
& 
& $\left(\overline{3},1,-2/3 \right)$
& $u_{\rm R}$
\\
\null
& $\left(\overline{4},1,-1/2\right)$
& $\left(1,1,2,1/2 \right)$
& $\left(1,1,1 \right)$
& $e_{\rm R}$
\\
&
& 
& $\left(1,1,0 \right)$
& $\nu_{\rm R}$
\\
\hline 
\end{tabular}
\caption{Decomposition of $16$ under PS, LQ, LR and SM symmetries with quantum numbers in an obvious notation. The last column denotes the name of the corresponding fermion.}
\label{tab:16decomp}
\end{table}

\textit{(ii) Gauge boson sector. } Besides the SM gauge bosons, we also have the color triplet, weak singlet PS leptoquark $X_{\rm PS}$ with the hypercharge ${\rm Y} = \pm 2/3$; the RH analogs of $W$ and $Z$ bosons, weak singlets $W_R$ and $Z_R$; and two SM weak doublet and color triplet heavy gauge bosons that mediate proton decay, $X,Y$ and $X',Y'$, with the hypercharges  
${\rm Y} = \pm 5/3$ and $ {\rm Y}' = \pm 1/3$, respectively. The 
12 $X,Y$ states complete the 24 gauge bosons of $SU(5)$, and the additional 21 states
$X',Y'$, $X_{\rm PS}$, $W_R$ and $Z_R$ complete the 45 gauge bosons of $SO(10)$. We give the decomposition of these gauge bosons in the last column of Table~\ref{tab:45decomp}. Their masses in terms of scalar VEVs can be found in Appendix~\ref{appendix:gaugemasses}. 
\renewcommand{\arraystretch}{1.2}
\begin{table}[h!]
\centering
\begin{tabular}{cccc|c}
\hline 
 $4_C\,2_L\,2_R $
& $4_C\,2_L\,1_R $
& $3_c\,2_L\,2_R\,1_{BL} $
& $3_c\,2_L\,1_Y $
&GBs
\\
\hline
 $\left({ 1,1,3} \right)$
& $\left({ 1,1},+1 \right)$
& $\left({ 1,1,3},0 \right)$
& $\left({ 1,1},+1 \right)$
&$W_{\rm R}^+$
\\
\null
& $\left({ 1,1},0 \right)$
&
& $\left({ 1,1},0 \right)$
& $W_{\rm R}^3$
\\
\null
& $\left({ 1,1},-1 \right)$
&
& $\left({ 1,1},-1 \right)$
& $W_{\rm R}^-$
\\
$\left({ 1,3,1} \right)$
& $\left({ 1,3},0 \right)$
& $\left({ 1,3,1},0 \right)$
& $\left({ 1,3},0 \right)$
& $W_{\rm L}^{\pm},W_{\rm L}^3$
\\
$\left({ 6,2,2} \right)$
& $\left({ 6,2},\pm\frac{1}{2} \right)$
& $\left({ 3,2,2},-\frac{1}{3} \right)$
& $\left({ 3,2},\frac{1}{6} \right)$
&$X,Y$
\\
\null
& 
&
& $\left({ 3,2},-\frac{5}{6} \right)$
&  $X',Y'$
\\
\null
&
& $\left({ \overline{3},2,2},+\frac{1}{3} \right)$
& $\left({ \overline{3},2},+\frac{5}{6} \right)$
&$\overline{X},\overline{Y}$
\\
\null
&
&
& $\left({ \overline{3},2},-\frac{1}{6} \right)$
&$\overline{X}',\overline{Y}'$
\\
$\left({ 15,1,1} \right)$
& $\left({ 15,1},0 \right)$
& $\left({ 1,1,1},0 \right)$
& $\left({ 1,1},0 \right)$
&$X_{\rm BL}$
\\
\null
&
& $\left({ 3,1,1},+\frac{2}{3} \right)$
& $\left({ 3,1},+\frac{2}{3} \right)$
&$\overline{X}_{\rm PS}$
\\
\null
&
& $\left({ \overline{3},1,1},-\frac{2}{3} \right)$
& $\left({ \overline{3},1},-\frac{2}{3} \right)$
& $X_{\rm PS}$
\\
\null
&
& $\left({ 8,1,1},0 \right)$
& $\left({ 8,1},0 \right)$
& $\text{gluons}$
\\
\hline 
\end{tabular}
\caption{Decomposition of $45$ under PS, LQ, LR and SM symmetries with quantum numbers in an obvious notation. To ease the reader's pain, the last column denotes the name of the $SO(10)$ gauge boson.}
\label{tab:45decomp}
\end{table}

\textit{(iii) Higgs scalar sector. } At the large scales we have $45_{\rm H}$ and $16_{\rm H}$ scalar representations. The decomposition of the former follows the one of the gauge bosons, with the provision that the heavy gauge bosons eat the scalar state with the same quantum number. The decomposition of the $16_{\rm H}$ is exactly the same as the one of $16_{\rm F}$ consisting of the SM fermions plus the right handed neutrino and is reported in Table~\ref{tab:16decomp}. 

\vspace{.3cm}
\paragraph*{\textbf{Single-step breaking. }}
For completeness we report here the unification condition for the single-step breaking at 1-loop. We also perform the one-loop RGE analysis for the three SM couplings by including the surviving scalar thresholds into their evolution. As we expect from the discussion, and verified numerically, the impact of $d=6$ gauge operators is small and unification is mainly fixed by scalar particle content instead. However, there are regimes, as we shall see, where their contribution becomes relevant.

Analitically, the unification conditions are
\begin{equation}
    \begin{split}
&\frac {M_{\rm GUT}} {M_{\rm Z}} \simeq \exp \left[{\frac{\pi}{20}\left(5\alpha_{1}^{-1}-3\alpha_{2}^{-1}-2\alpha_{3}^{-1}\right)}\right]
\\
&\qquad\qquad \qquad\qquad\qquad\cdot \left[\left(\frac{M_{\rm Z}^4}{m_3m_8 m_{ \tilde q}^2}\right)\right]^{\frac{1}{40}} \,,\\
&\frac {M_{\rm GUT}} {M_{\rm Z}} \simeq \exp \left[{\frac{6\pi}{13}\left(\alpha_{2}^{-1}-\alpha_{3}^{-1}\right)}\right]
\\
&\qquad\qquad\qquad\qquad\qquad\cdot \left[\left(\frac{m_3^2m_{\tilde{\ell}}^2 m_{ \tilde q}}{m_8^3 M_{\rm Z}^2}\right)\right]^{\frac{1}{26}},
\label{mainconstraint}
\end{split}
\end{equation}
where the above $\alpha_i$ are evaluated at the scale $M_{\rm Z}$, $m_3, m_8, m_{\tilde q}, m_{\tilde{\ell}}$ denote the masses of the weak triplet, the color octet, the quark doublet and the lepton doublet-like scalars respectively (the last one denotes actually the average mass of two states from $45_{\rm H}$ and $16_{\rm H}$). The choice of the first condition has the advantage of being independent of the weak doublets mass scale - a useful choice, given that the numerical results seem to be insensible to it. No other combination of gauge couplings allows for cancellation of mass thresholds, therefore we chose to give the condition for the $2-3$ gauge coupling.

For $\alpha_1 = 59.02^{-1}$, $\alpha_2 = 29.57^{-1}$, $\alpha_3 = 8.44^{-1}$, and $M_{\rm Z} = 91.19\,\rm GeV$, the above conditions are a good approximation of the two loop result. 
In fact, for $m_3\simeq m_8 \simeq m_{\tilde{q}} \simeq M_{\rm GUT}$ we obtain $M_{\rm GUT}\simeq 5\cdot 10^{13}\rm GeV$ from the first equation, while we get $M_{\rm GUT}\simeq 2\cdot 10^{16}\rm GeV$.

This ought to be fixed by taking light particles other than the color octet, as the second Eq.~\eqref{mainconstraint} seems to informs us of (notice also that the scalars have less impact in the first condition, enforcing this requirement). In fact, by performing the one-loop RGE analysis for this scenario we find that a light color octet pushes the weak triplet, quark doublet and lepton-like doublet to be light as well. For $m_8\simeq 10-100 {\rm TeV}$, we necessarily need $m_3\simeq m_{\tilde{q}}\simeq m_{\tilde{\ell}}\sim {\rm TeV}$ to achieve unification. Additionally, the inclusion of the $d=6$ corrections to the gauge couplings, of the order of $\mathcal{O}(10^{-2})$, is also required. In particular, these higher-dimensional corrections become increasingly important for lower values of the color octet mass. In this regime, when $m_8\leq 100{\rm TeV}$, unification can be achieved only in their presence.

Moreover, for $m_8\simeq 10^{8}\rm GeV$, and $m_3\simeq m_{\tilde{\ell}}\simeq m_{\tilde{q}}\simeq \rm TeV$, we obtain, for both conditions in \eqref{mainconstraint}, $M_{\rm GUT}\simeq 5\cdot 10^{14}\rm GeV$, which is analogous to what we found in the two loop RG analysis discussed below. 

For such value of unification -- and $\langle 16_{\rm H}\rangle$ -- the neutrino mass scale is found to be larger than about $0.1\,$eV, c.f. Fig.~\ref{fig:numass}. Correspondingly, the color octet, the weak triplet and the quark-like scalar need to be light. However, things are not so simple: relaxing the degeneracy of the VEVs leads to the loss of the above correlations. However not all is lost, as we are about to verify from in the full analysis. 

\vspace{0.3cm}
\paragraph*{\textbf{Full analysis.}}

\begin{figure*}[th!]
\centering
    \includegraphics[width=.99\textwidth]{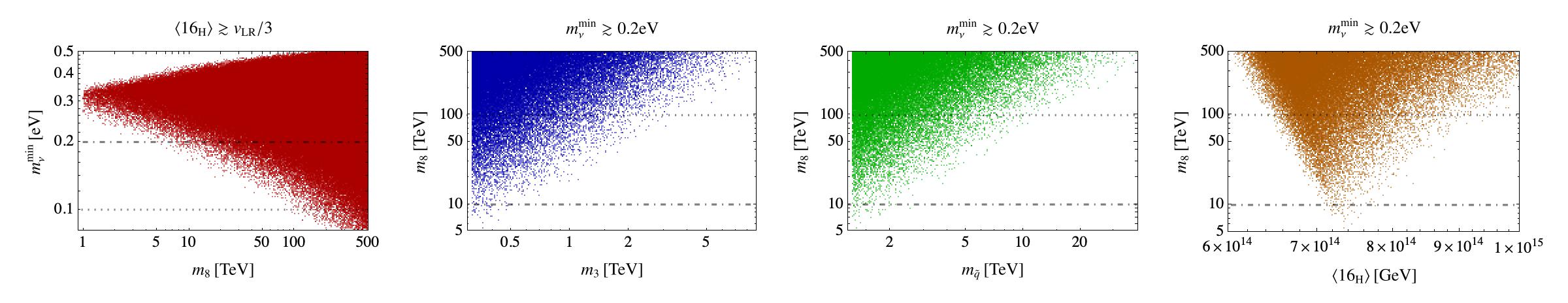}
    \caption{Gauge coupling unification realizations for different values of thresholds masses. The first panel shown $m_{8}$ scale as a function of the minimum possible neutrino mass $m_{\nu}^{\rm min}$ according to Eq.~\eqref{nueffect}. The second and third panel shows the impact of the color octet on the weak triplet scale $m_3$ and the scalar quark $m_{\rm sq}$ respectively. The final panel shows the resulting $\langle 16_{\rm H}\rangle$ scale dependence on $m_{8}$. The horizontal lines denote example values of neutrino minimal masses in the first panel. In the second panel just signal the $m_8 \simeq 10 - 100 \rm TeV$ benchmark values. The last three panels assume $m_{\nu}\lesssim 0.2 \rm eV$, while the first panel is obtained under the requirement $\langle 16_{\rm H}\rangle \gtrsim v_{\rm LR}/3$ -- see main text for more details. 
  }
    \label{fig:thresholds}
\end{figure*}

Armed with the knowledge from the previous Subsection we now fully tackle the gauge couple unification analysis by performing a Monte-Carlo analysis of the two-loop RG flow. First, we fix the ratio between $v_{\rm QL}$ and $v_{\rm LR}$ and then we randomize particle thresholds scales. To fulfill unification conditions, we find the appropriate value of $\delta \alpha_i$ in \eqref{eq:deltaalphai}. If $M_{\rm GUT}/\Lambda \gtrsim 10^{-1}$, the found realization is excluded from the sampling. 

Due to the closeness of symmetry breaking scales, gauge boson mass-thresholds are split around the unification scale. Since we are performing a two-loop analysis, proper matching condition requires correction of the SM gauge couplings $\alpha_{i}$ ($i=1,2,3$) as given in~\cite{Hall:1980kf,Weinberg:1980wa}
\begin{equation}
\label{lambdaigeneral}
\begin{split}
   & \frac{1}{\alpha_{i}(\mu)} = \frac{1}{\alpha_{\rm G}(\mu)} - \lambda_i(\mu)\,,\\
    & \lambda_{i}(\mu) = \frac{1}{12\pi}\left({\rm Tr}\,t_{\rm V}^2 - 21 {\rm Tr}\,t_{\rm V}^2\log\left[\frac{M_{\rm V}}{\mu}\right] \right)\,,
    \end{split}
\end{equation}
where $\lambda_i(\mu)$ are obtained from the effective gauge theory description derived integrating out the heavy particles, $t_{\rm V}$ correspond to the gauge boson representation generators and $M_{\rm V}$ their mass thresholds. We neglect here contributions from the scalar sector as they enter with a suppressed numerical factor, and the fermionic contribution vanishes since there are no new fermions at the GUT scale. 

Normally, one chooses $\mu= M_{\rm V}$ such that the scale dependent contribution in \eqref{lambdaigeneral} drops out. However, since in our case gauge boson masses are split, c.f.,~\eqref{eq:gaugemasses}, such a simple choice is not available. 

In fact, assuming $|v_{\rm LR}|\neq |v_{\rm QL}|$, we obtain
\begin{equation}
    \begin{split}
        &\lambda_3(\mu) = \frac{1}{6 \pi}\left\{\frac{5}{2} - 21 \left[ \frac{1}{2}\log \frac{m_{X_{\rm PS}}}{\mu} + \log \frac{m_{ X}}{\mu} + \log \frac{m_{ X'}}{\mu}   \right] \right\}\,,\\
    &\lambda_2(\mu) = \frac{1}{6 \pi}\left\{3- 21 \left[ \frac{3}{2}\log \frac{m_{ X}}{\mu} + \frac{3}{2}\log \frac{m_{ X'}}{\mu}  \right] \right\}\,,\\
        &\lambda_1(\mu) = \frac{1}{5 \pi}\left\{\frac{20}{3}- 21 \left[ \log \frac{m_{ W_{\rm R}}}{\mu} + \frac{4}{3}\log \frac{m_{X_{\rm PS}}}{\mu} +\right. \right. \\
        &\qquad \qquad\qquad \qquad\qquad \left. \left. \frac{25}{6}\log \frac{m_{ X}}{\mu} + \frac{1}{6}\log \frac{m_{ X'}}{\mu}  \right] \right\}\,.
    \end{split}
\end{equation}
Choosing $\overline{\mu} = \sqrt{m_{ X}m_{X'}}$, significant simplification takes place
\begin{equation}
    \begin{split}
        &\lambda_3(\overline{\mu}) = \frac{1}{6 \pi}\left\{\frac{5}{2} - \frac{21}{2} \log \frac{m_{X_{\rm PS}}}{\overline{\mu}} \right\}\,,\\
        &\lambda_2(\overline{\mu}) = \frac{1}{2 \pi}\,,\\
        &\lambda_1(\overline{\mu}) = \frac{1}{5 \pi}\left\{\frac{20}{3}- 21 \left[ \log \frac{m_{ W_{\rm R}}}{\overline{\mu}} + \frac{4}{3}\log \frac{m_{X_{\rm PS}}}{\overline{\mu}} + 4\frac{m_{ X}}{\overline{\mu}}  \right] \right\}\,.
    \end{split}
\end{equation}
Furthermore, scalar thresholds around $M_{\rm GUT}$ have been varied using conditions analogous to the one above. No significant alterations of our results have been noticed. 

As expected, regardless of the ratio $v_{\rm QL}/v_{\rm LR}$, neutrino masses down to $0.1\,$eV can be obtained with rather arbitrary scalar thresholds scales. The typical scale of $M_{X,X'}$ gauge bosons, responsible for proton decay, also varies between $10^{14} - 10^{15}\,$GeV, therefore requiring some amount of cancellation in proton decay amplitudes. However, 
the theory is far from being ruled out in foreseeable future~\cite{Super-Kamiokande:2020wjk,DUNE:2020lwj,JUNO:2015zny}, since, as found by~\cite{Senjanovic:2024uzn}, $M_{X}\simeq 8 \cdot 10^{13}\,\rm GeV$, is still compatible with current experimental data.
 
However, an interesting correlation emerges, as summarized by Fig.~\ref{fig:thresholds}. The figure shows realizations of GUT as a function of the octet mass $m_8$. Specifically, $v_{\rm LR}\simeq 3 v_{\rm QL}$ is chosen to account for the realistic fermion spectrum. Note that this implies $\langle16_{\rm H}\rangle\sim v_{\rm QL}$, as discussed in the Yukawa analysis of the previous Section.

\subsection{Phenomenological implications}
From the figures the following correlations emerge: A color octet below about $5\,\rm TeV$ ($100\,$TeV) implies $m_{\nu}\gtrsim 0.2\,$eV ($0.1\,$eV), $m_{3}\lesssim 500\,\rm GeV$ ($5\,$TeV) and $m_{\tilde q}\lesssim 3\,$TeV ($10\,$TeV). Finally, $\langle 16_{\rm H}\rangle\simeq 7-8 \cdot 10^{14}\,\rm GeV$ ($7-9 \cdot 10^{14}\,$GeV). Similarly to the findings of \cite{Preda:2022izo}, the weak doublet mass $m_{\ell}$ is rather unconstrained and the other particle thresholds live near $M_{\rm GUT}$. 

A realistic fermion mass spectrum can also be obtained by choosing $v_{\rm LR}<v_{\rm QL}$, which leads to milder correlations. This can be understood from the fact that, when $v_{\rm LR}\ll v_{\rm QL}$, the $\langle 16_{\rm H}\rangle$ needs to be lighter to ensure unification, therefore leading to larger values of neutrino masses for similar scales of the color octet. 
Similar features were noted in~\cite{Preda:2022izo} and verified numerically. Therefore, the case shown in Fig.~\ref{fig:thresholds} provides the absolute lower limit on the neutrino mass given by a potentially light color octet. Notice that in the analysis, also other scalar thresholds were included. Moving them significantly away from $M_{\rm GUT}$ simply strengthens our findings.

Eventually, KATRIN will probe neutrino mass down to $0.2\,$eV~\cite{Katrin:2024tvg}. On the other hand, for this to be a strict prediction of the model, a light color octet - below $10\, \rm TeV$ - needs to be found at colliders. 

The phenomenological implications of the new light scalar states were addressed in~\cite{Preda:2022izo}. Of particular interest is the weak triplet, since its small VEV naturally induces a small deviation for the Standard Model value of $W-$boson mass~\cite{Senjanovic:2022zwy,Senjanovic:2023jvv}.

\section{Conclusions and outlook}

The present work is a continuation of a recent study~\cite{Preda:2022izo} on the minimal $SO(10)$ GUT theory. The theory is based on the following Higgs representations: $16_{\rm H}$ and $45_{\rm H}$, responsible for the high-scale symmetry breaking, and complex $10_{\rm H}$ responsible for the electroweak scale symmetry breaking. In order for the theory to be realistic, a renormalizable model must be augmented by high-dimensional operators, which ensure correct fermion mass relation, non-vanishing neutrino mass and successful gauge coupling unification. 

Ref.~\cite{Preda:2022izo} showed that, in the absence of flavour rotation in proton decay amplitudes, consistency of the theory would predict the following light scalar states at present-day or near future collider energies: a weak triplet, a color octet and a weak doublet, color triplet. Strictly speaking, however, a complete analysis must allow for the flavour freedom of the theory which admittedly requires some amount of fine tuning. Although practitioners of the field tend to shy away from this freedom, we believe that a complete analysis is a must. Here we addressed precisely this issue and thus completed the work of~\cite{Preda:2022izo}.

As expected, the predictions of~\cite{Preda:2022izo} are relaxed: The new scalar states might not be accessible at colliders in any foreseeable future. However, not all is lost. We find that a light color octet, with a mass below $5\,\rm TeV$, implies a series of interesting correlations. First of all, it implies a lower bound on neutrino mass of order $0.2\,\rm eV$, potentially within the reach of KATRIN experiment~\cite{Katrin:2024tvg} in the foreseeable future. Moreover, the weak triplet would lie below $500\,\rm GeV$, tantalizingly close to being accessible at the LHC,
and the weak doublet, color triplet below $3\,\rm TeV$. In other words, a light color octet would recover the rest of the findings of~\cite{Preda:2022izo}.

It is instructive to compare the theory with the minimal effective $SU(5)$ model analyzed recently~\cite{Senjanovic:2024uzn}. Both need higher-dimensional operators, but there is an essential difference: The leading $d=5$ operator in the gauge kinetic sector vanishes in the $SO(10)$ model, thus becoming effectively negligible. In turn, this forces the above scalar state to populate the desert in energies between $M_{\rm GUT}$ and $M_{W}$ in order to achieve unification of gauge couplings. 

Second, unlike in $SU(5)$, $SO(10)$ is tailor-made for a natural generation of neutrino mass. Nevertheless, the choice of minimal scalar representations still forces higher-dimensional operator to source the right-handed neutrino mass. In turn, this pushes the theory into single-step breaking. 

When the dust settles, though, a serious problem remains, plaguing basically all realistic grand unified theories: the lack of predictions for nucleon decay branching ratios. This jeopardises the very idea of unification, for nucleon decay lies at its heart. The essence of the problem is a huge scale of unification, totally out of experimental reach. However, this tells that one could go effective all the way and try the expansion in a small parameter $M_W/M_{\rm GUT}$.
In his classic work~\cite{Weinberg:1979sa,Senjanovic:2009kr}, Weinberg found interesting consequences of nucleon decay being driven by high-scale physics, which could serve as a serious test of the grand unification idea. One of the predictions is remarkable: Neutron has no two body decay into a Kaon and charged lepton.

{\bf Acknowledgments} We are grateful to Borut Bajc, Gia Dvali, Alejandra Melfo and Vladimir Tello for illuminating discussions. We further thank Uros Seljak and Zvonimir Vlah for useful comments on cosmological neutrino-mass bound. 
MZ acknowledges support by the National Natural Science Foundation of China (NSFC) through the grant No.\ 12350610240 ``Astrophysical Axion Laboratories''

\bibliography{biblio}

\appendix
\section{Breaking patterns}
\label{app:masses}

Under \eqref{eq:45ansatzvLRQL}, the potential for the vacuum becomes
 \begin{equation}
 \begin{split}
&V_0= -m^2 \left(2 {v_{\rm LR}}^2+3 {v_{\rm QL}}^2\right)+{\lambda_1} \left(2 {v_{\rm LR}}^2+3 {v_{\rm QL}}^2\right)^2\\
&+{\lambda_2} \left(2
   {v_{\rm LR}}^4+3 {v_{\rm QL}}^4\right)+ \frac{\left(2 {v_{\rm LR}}^2+3
   {v_{\rm QL}}^2\right)^3}{{\mu_2^2}}
   \\& +
   \frac{2 {v_{\rm LR}}^6+3 {v_{\rm QL}}^6}{{\mu_3^2}}+
   \frac{\left(2 {v_{\rm LR}}^4+3 {v_{\rm QL}}^4\right) \left(2 {v_{\rm LR}}^2+3
   {v_{\rm QL}}^2\right)}{{\mu_1^2}},
   \end{split}
 \end{equation}
leading to the extremization conditions
 \begin{equation}
 \label{eq:pdv_LRV}
 \begin{split}
 	&\partial_{v_{\rm LR}}V_0 = 4 {v_{\rm LR}} \left(-m^2+2 {v_{\rm LR}}^2 (2 {\lambda_1}+{\lambda_2})+6 {\lambda_1} {v_{\rm QL}}^2\right)\\&+\frac{12
   {v_{\rm LR}}^5}{{\mu_3^2}}
   +\frac{4 {v_{\rm LR}} \left(6 {v_{\rm LR}}^4+6 {v_{\rm LR}}^2 {v_{\rm QL}}^2+3 {v_{\rm QL}}^4\right)}{{\mu_1^2}}\\&+
   \frac{4 {v_{\rm LR}} \left(12 {v_{\rm LR}}^4+36 {v_{\rm LR}}^2 {v_{\rm QL}}^2+27 {v_{\rm QL}}^4\right)}{{\mu_2^2}}\,,
   \end{split}
 \end{equation}
 \begin{equation}
  \label{eq:pdv_QLV}
 \begin{split}
 	&\partial_{v_{\rm QL}}V_0= 6 {v_{\rm QL}} \left(-m^2+4 {\lambda_1} {v_{\rm LR}}^2+2 {v_{\rm QL}}^2 (3 {\lambda_1}+{\lambda_2})\right)
  \\&
 + \frac{6
   {v_{\rm QL}} \left(2 {v_{\rm LR}}^4+4 {v_{\rm LR}}^2 {v_{\rm QL}}^2+9 {v_{\rm QL}}^4\right)}{{\mu_1^2}}+\frac{18 {v_{\rm QL}}^5}{{\mu_3^2}}
   \\&
   +
   \frac{6 {v_{\rm QL}} \left(12
   {v_{\rm LR}}^4+36 {v_{\rm LR}}^2 {v_{\rm QL}}^2+27 {v_{\rm QL}}^4\right)}{{\mu_2^2}}\,.
   \end{split}
	 \end{equation}
In the case where both $v_{\rm LR}$ and $v_{\rm QL}$ are non-vanishing and $|v_{\rm LR}|\neq |v_{\rm QL}|$, a useful combination between \eqref{eq:pdv_LRV} and \eqref{eq:pdv_QLV} is given by
\begin{equation}
\begin{split}
 \label{eq:pdv_LRv_QLv_QL}
	&\frac{1}{v_{\rm LR}^2-v_{\rm QL}^2}\left(v_{\rm QL} \partial_{v_{\rm LR}} - \frac{2}{3}v_{\rm LR}\partial_{v_{\rm QL}}\right)V_0=\\&+8 {\lambda_2} {v_{\rm LR}} {v_{\rm QL}}+ \frac{4 {v_{\rm LR}} {v_{\rm QL}} \left(4 {v_{\rm LR}}^2+6 {v_{\rm QL}}^2\right)}{{\mu_1}^2}\\&
 +\frac{4 {v_{\rm LR}} {v_{\rm QL}} \left(3
   {v_{\rm LR}}^2+3 {v_{\rm QL}}^2\right)}{{\mu_3}^2}.
   \end{split}
   \end{equation}
   Eq.~\eqref{eq:pdv_LRv_QLv_QL} or \eqref{eq:pdv_QLV} combined with \eqref{eq:pdv_LRv_QLv_QL} provide a replacement for $m^2$ and $\lambda_2$ respectively. 

 For completeness, although we already eliminated the full symmetry brreaking pattern $v_{\rm LR}\sim v_{\rm QL}\neq 0$,  we report also on the mass matrix $m^2_{AC}$ of the $A$ and $C$ state, which is non-diagonal
   \begin{equation}
 \label{eq:45total}
 \begin{split}
 		&\langle45_{\rm H}\rangle  = i \, \sigma_2 \otimes {\rm{diag}}(v_{\rm LR}+A  ,v_{\rm LR}+A,\\
   &\qquad \qquad \qquad \qquad v_{\rm QL}+C,v_{\rm QL}+C,v_{\rm QL})\,,\\
 	&(m_{AC}^2)_{11} =32 {\lambda_1} {v_{\rm LR}}^2+\frac{8
   \left(24 {v_{\rm LR}}^4+36 {v_{\rm LR}}^2 {v_{\rm QL}}^2\right)}{{\mu_2}^2}
  \\&
  \qquad \qquad + \frac{64 {v_{\rm LR}}^4}{{\mu_1}^2}
  +\frac{8
   \left(3 {v_{\rm LR}}^4-3 {v_{\rm LR}}^2 {v_{\rm QL}}^2\right)}{{\mu_3}^2}\,,\\
 	&(m_{AC}^2)_{22}=32 {\lambda_1} {v_{\rm QL}}^2+\frac{8 {v_{\rm QL}}^2 \left(24 {v_{\rm LR}}^2+36 {v_{\rm QL}}^2\right)}{{\mu_2}^2}\\
  &\qquad \qquad+\frac{8 {v_{\rm QL}}^2 \left(3 {v_{\rm QL}}^2-3 {v_{\rm LR}}^2\right)}{{\mu_3}^2}
  +\frac{64 {v_{\rm QL}}^4}{{\mu_1}^2}\,,\\
 	&(m_{AC}^2)_{12} =32 {\lambda_1} {v_{\rm LR}} {v_{\rm QL}}+\frac{32 {v_{\rm LR}} {v_{\rm QL}} \left({v_{\rm LR}}^2+{v_{\rm QL}}^2\right)}{{\mu_1}^2}\\
  &\qquad \qquad
  +\frac{32 {v_{\rm LR}} {v_{\rm QL}} \left(6
   {v_{\rm LR}}^2+9 {v_{\rm QL}}^2\right)}{{\mu_2}^2}.
   \end{split}
 \end{equation}
 The determinant of the $AC$ mass matrix is given by
 \begin{equation}
 	-\frac{64 {v_{\rm LR}}^2 {v_{\rm QL}}^2 (3 {\mu_1}^2+4 {\mu_3}^2)^2 \left({v_{\rm LR}}^2-{v_{\rm QL}}^2\right)^2}{{\mu_1}^4
   {\mu_3}^4},
 \end{equation}
 which is clearly non-positive - yet another tachyon. In this case, however, one could in principle trade the tachyon for the massless scalar. Namely, since in the case of interest $v_{\rm LR}\neq v_{\rm QL}$, this could be achieved by requiring the following (obviously unstable) condition
 \begin{equation}
 \label{eq:mu1mu3rel}
 3\mu_1^2+ 4 \mu_3^2 =0.
 \end{equation}
 Under this condition, the positivity of the trace would impose
 \begin{equation}
 	4 {\mu_3}^2 \left({\lambda_1} {\mu_2}^2+6 {v_{\rm LR}}^2+9 {v_{\rm QL}}^2\right)-3 {\mu_2}^2
   \left({v_{\rm LR}}^2+{v_{\rm QL}}^2\right)>0
 \end{equation}
 which would admit solution for $\mu_2^2$ and $\mu_3^2$ positives, with $\mu_2^2$ big enough to guarantee the asymtptotic boundedness of the potential. Finally, the condition \eqref{eq:mu1mu3rel}, combined with \eqref{eq:pdv_LRv_QLv_QL}, would give $\lambda_2 = 3 v_{\rm QL}^2/\mu_3^2$. However, all of this is just of academic interest, since as shown above, the theory has the tachyon, period. Just as in the radiative Coleman-Weinberg case, we are forced to look at partial symmetry breaking patterns.
 
 \subsection{Mass spectrum for QL and LR breaking pattern }\label{QL_LR_mass}
We report here the resulting mass spectrum of the $LR$ and $QL$ breaking patterns taking into account the inclusion of higher-dimensional operators. 
 The former direction leads to
 \begin{equation}
 	m_3^2 = 8 \lambda_2{v_{\rm LR}}^2+8{v_{\rm LR}}^4 \left(\frac{2}{{\mu_1}^2}+\frac{3}{{\mu_3}^2}\right),
 \end{equation}
 \begin{equation}
 	m_{15}^2=-4
   {\lambda_2} {v_{\rm LR}}^2+{v_{\rm LR}}^4 \left(-\frac{8}{{\mu_1}^2}-\frac{6}{{\mu_3}^2}\right),
 \end{equation}
 \begin{equation}
 	m_1^2=8 {v_{\rm LR}}^2 \left(2 {\lambda_1}+{\lambda_2}+3 {v_{\rm LR}}^2
   \left(\frac{2}{{\mu_1}^2}+\frac{4}{{\mu_2}^2}+\frac{1}{{\mu_3}^2}\right)\right).
 \end{equation}
 The first two equations denote the masses of the weak triplet and of the adjoint $SU(4)_C$ scalar fields respectively, while $m_1$ denotes the mass of the associated Higgs boson in this case, the QL singlet. Notice that at the renormalizable level the presence of tachyonicity forbids the symmetry breaking in this channel. However, this is easily cured by the $d=6$ terms. 

 A similar situation emerges in the $QL$ breaking pattern, in which case the masses are given by
 \begin{equation}
 	m_{\rm 3L}^2=-4 {\lambda_2} {v_{\rm QL}}^2+m_{\rm 3R}^2 = -\frac{6 {v_{\rm QL}}^4 \left({\mu_1}^2+2 {\mu_3}^2\right)}{\left({\mu_1} {\mu_3}\right)^2},
 \end{equation}
 \begin{equation}
 	m_8^2=8\lambda_2 {v_{\rm QL}}^2 +\frac{24 {v_{\rm QL}}^2 \left({\mu_1}^2+{\mu_3}^2\right)}{\left({\mu_1} {\mu_3}\right)^2},
 \end{equation}
 \begin{equation}
 	m_{\rm X}^2=8 {v_{\rm QL}}^2 \left(3 {\lambda_1}+{\lambda_2}+3 {v_{\rm QL}}^2
   \left(\frac{3}{{\mu_1}^2}+\frac{9}{{\mu_2}^2}+\frac{1}{{\mu_3}^2}\right)\right)\,,
 \end{equation}
where $m_{3L,R}$ denotes the masses of the left right triplets, $m_8$ the color octet mass and $ m_X$ the mass of the Higgs boson. Once again, also here, the renormalizable level tachyon can easily be cured by the $d=6$ operators, therefore making this breaking pattern viable. In other words, since the theory requires in any case $d=5$ and $d=6$ terms, there is no need for radiative corrections. 

 \section{Gauge boson masses from $45_{H}$ and $16_{\rm H}$}\label{appendix:gaugemasses}
 Adopting the notation and convention of Ref.~\cite{Bertolini:2009es}, it is straightforward to compute the contribution to the gauge boson masses due to $\langle 45_{\rm H}\rangle$ and $\langle 16_{\rm H}\rangle= \tilde{\nu}_{\rm R}$
    \begin{equation} 
    \label{eq:gaugemasses}
    \begin{split}
 	& m^2(1_{\rm C},1_{\rm L},1_{\rm Y}) =m_{W_R}^2=4 g^2 v_{\rm LR}^2 + g^2 \tilde{\nu}_{\rm R}^2\,,
  \\&m^2(\overline{3}_{\rm C},1_{\rm L},2/3_{\rm Y}) =m_{X_{\rm PS}}^2=4 g^2 v_{\rm QL}^2 + g^2 \tilde{\nu}_{\rm R}^2\,,
  \\&m^2(1_{\rm C},3_{\rm L},0_{\rm Y}) =0\,,
  \\&m^2(8_{\rm C},1_{\rm L},0_{\rm Y}) =0\,,
  \\&m^2(3_{\rm C},2_{\rm L},-5/6_{\rm Y}) =m_{X}^2= g^2(v_{\rm QL} - v_{\rm LR})^2\,,
  \\&m^2(3_{\rm C},2_{\rm L},1/6_{\rm Y}) =m_{ X'}^2 =g^2(v_{\rm QL} + v_{\rm LR})^2 + g^2 \tilde{\nu}_{\rm R}^2 \,\,
  \\&m^2(1_{\rm C},1_{\rm L},0_{\rm Y}) =\begin{pmatrix}
3/2 &\sqrt{3/2}\\
\sqrt{3/2} &1
\end{pmatrix}g^2 \tilde{\nu}_{\rm R}^2\,
   \end{split}
 \end{equation}
 where the number in parenthesis denote the multiplet charge under Standard Model symmetry group $3_{\rm C}2_{\rm L}1_{\rm Y}$. The last term correspond to the $2\times 2$ mass matrix of $W_{\rm R}^3$, $X_{\rm BL}$, c.f., Table~\ref{tab:45decomp}.

\section{Gauge coupling unification coefficients}\label{app:deltaalphai}
Taking into account the VEV structure, given by eq.~\eqref{vev}, and the fact that only the higher-dimensional operator
\begin{equation}
    {\rm Tr}\left(F_{\mu\nu}F^{\mu\nu} \langle 45_{\rm H}^2\rangle  \right)/\Lambda^2
\end{equation}
modifies the coupling unification condition, we can deduce the numerical coefficients correcting the couplings at the unification scale. To that end, is useful to write the generators of the gauge groups in terms of the $SO(10)$ Cartan subalgebra $\{\Sigma_{12}, \Sigma_{34},\cdots, \Sigma_{910}\}$:
\begin{align}
T_{3{\rm L}}&=\frac{1}{2}(\Sigma_{78}+\Sigma_{910})\, ,\\
\nonumber
T_{3{\rm R}}&=\frac{1}{2}(\Sigma_{78}-\Sigma_{910})\, ,\\
\nonumber
T_{3{\rm C}}&=\frac{1}{2}(\Sigma_{12}-\Sigma_{34})\, ,\\
\nonumber
{ B-L}&=-\frac{2}{3}(\Sigma_{12}+\Sigma_{34}+\Sigma_{56})\, ,
\end{align}
where the indices $\{12,34,56\}$ denote the $SO(6)\simeq SU(4)$ subspace, while the indices $\{78,910\}$ denote the $SO(4)\simeq SU(2)\times SU(2)$ subspace.

For the LR case, we see that the VEV of $45_{\rm H}$ lies in the $SO(6)$ direction and therefore the terms ${\rm Tr}\:(T_{3{\rm L},3{\rm R}}^{2}\langle 45_{\rm H}\rangle_{\rm LR}^{2})$ vanish. This in turn implies no correction to $\alpha_{2}$ coupling. On the other hand, the correction to $\alpha_{3}$ is non-zero due to the following term:
\begin{equation}
{\rm Tr}\:(T_{3{\rm C}}^{2}\langle 45_{\rm H}\rangle_{\rm LR}^{2})=\frac{1}{4}{\rm Tr}\:\left((\Sigma_{12}-\Sigma_{34})^{2}\langle 45_{\rm H}\rangle_{\rm LR}^{2}\right)=v_{\rm LR}^{2}.
\end{equation}
For the $\alpha_{1}$ correction, we know that:
\begin{equation}
\frac{ Y}{2}=T_{3{\rm R}}+\frac{ B-L}{2}.
\end{equation}
The $U(1)$ generator is normalized as $T_{1}=\sqrt{\frac{3}{5}}\frac{ Y}{2}=\sqrt{\frac{3}{5}}\left(T_{3{\rm  R}}+\frac{B-L}{2}\right)$, resulting in:
\begin{equation}
\label{eq:T1}
T_{1}=\sqrt{\frac{3}{5}}\left[\frac{1}{2}(\Sigma_{78}-\Sigma_{910})-\left(\frac{1}{3}\right)\left(\Sigma_{12}+\Sigma_{34}+\Sigma_{56}\right)\right]\, .
\end{equation}
We then find the $\alpha_1$ correction to be:
\begin{equation}
{\rm Tr}\left(T_{1}^{2}\langle 45_{\rm H}\rangle_{\rm LR}^{2}\right)=\frac{2}{5} v_{\rm LR}^{2}.
\end{equation}
The correction for the SM couplings at the unification scale in the case of LR intermediate symmetry:
\begin{equation}
\delta\alpha_{1}^{\rm LR}=\frac{2}{5}\left(\frac{v_{\rm LR}}{\Lambda}\right)^{2}\alpha_{1}\:,\:\:\:\ \delta\alpha_{2}^{\rm LR}=0\:,\: \:\:\: \delta\alpha_{3}^{\rm LR}=\left(\frac{v_{\rm LR}}{\Lambda}\right)^{2}\alpha_{3}.
\end{equation} 

For the QL case, the computation follows in a similar manner. The VEV is now in the $SO(4)$ direction, such that the corrections  at $M_{\rm GUT}$ will only impact $\alpha_{2}$ and $\alpha_{1}$. For the $\alpha_{2}$ coupling, the correction comes from:
\begin{equation}
{\rm Tr}\:(T_{3{\rm L}}^{2}\langle 45_{\rm H}\rangle_{\rm QL}^{2})=\frac{1}{4}{\rm Tr}\:\left((\Sigma_{78}+\Sigma_{910})^{2}\langle 45_{\rm H}\rangle_{\rm QL}^{2}\right)=v_{\rm QL}^{2}\, ,
\end{equation}
while for $\alpha_{1}$ it is given by:
\begin{equation}
\begin{split}
{\rm Tr}\:(T_{1}^{2}\langle 45_{\rm H}\rangle_{\rm QL}^{2})=\frac{3}{20}\left({\rm Tr}\:\left((\Sigma_{78}-\Sigma_{910})^{2}\right.\right.\\
\left.
\times \langle 45_{\rm H}\rangle_{\rm QL}^{2}\right)=\frac{3}{5} v_{\rm QL}^{2}.
\end{split}
\end{equation}
We used the same definition for the $U(1)$ generator as before to get the following corrections for the couplings at $M_{\rm GUT}$:
\begin{equation}
\delta\alpha_{1}^{\rm QL}=\frac{3}{5}\left(\frac{v_{\rm QL}}{\Lambda}\right)^{2}\alpha_{1}\:,\:\:\ \delta\alpha_{2}^{\rm QL}=\left(\frac{v_{\rm QL}}{\Lambda}\right)^{2}\alpha_{2} \:,\:\:\
\delta\alpha_{3}^{\rm QL}=0.
\end{equation}

\end{document}